\documentclass[11pt,twoside]{article}

\usepackage{asp2004}
\usepackage{epsf}
\usepackage{psfig}
\usepackage{lscape}
\usepackage{graphicx}
\usepackage{subfigure}

\begin{document}
\title{Near-Infrared Photometric Analyses of White Dwarf Stars}   
\author{P.-E. Tremblay, and P. Bergeron}   
\affil{D\'epartement de Physique, Universit\'e de Montr\'eal, C.P. 6128, Succ. 
Centre-Ville, Montr\'eal, Qu\'ebec, Canada, H3C 3J7} 

\begin{abstract} 
We review the available near- and mid- infrared photometry data sets
for white dwarfs from the Two Micron All-Sky Survey (2MASS) Point
Source Catalog and the Spitzer Space Telescope. These data sets have
been widely used to search for white dwarfs with an infrared excess as
well as to characterize the atmosphere of cool white dwarfs. We
evaluate the reliability of the 2MASS photometry by performing a
statistical comparison with published JHK CIT magnitudes, and by
carrying out a detailed model atmosphere analysis of the available
photometry. We then present a critical examination of various results
published in the literature including data from the Spitzer Space
Telescope.
\end{abstract}



\section{Introduction}

With the recent All-Sky Data Release of the Two Micron All-Sky
Survey\footnote{See
http://www.ipac.caltech.edu/2mass/releases/allsky/} (2MASS), we are
now able to retrieve near-infrared (NIR) $J$, $H$, and $K_S$
magnitudes for more than a thousand white dwarfs that fall within the
2MASS detection limit. This database was used in several studies aimed
at identifying new cool white dwarfs or circumstellar disks
\citep{kilic06} and seeking binary candidates
\citep{wachter03}. In addition to the 2MASS NIR photometry, there is a developing
interest to observe white dwarfs at longer wavelengths in the mid-infrared (MIR). The
{\it Spitzer Space Telescope} IRAC\footnote{See
http://ssc.spitzer.caltech.edu/irac/} photometry has been used in
recent surveys of relatively bright, nearby white dwarfs to better
constrain the atmospheric parameters of cool white dwarfs
\citep{kilic06b} and to seek MIR excesses from disks
\citep{hansen06}. Before undertaking a more systematic search of white dwarf stars in
binaries or of circumstellar disk systems using 2MASS or {\it Spitzer}
data, it seems appropriate as a first step to evaluate properly the
reliability of the infrared photometric data sets.

\section{Comparison of CIT and 2MASS Photometry}

Our photometric sample used to compare against the 2MASS data is drawn
from the detailed photometric and spectroscopic analyses of
\citet[][hereafter BRL97]{bergeron97} and \citet[][hereafter BLR01]{bergeron01} who obtained improved
atmospheric parameters of cool white dwarfs from a comparison of
optical $BVRI$ and infrared $JHK$ photometry with the predictions of
model atmospheres appropriate for these stars. We selected from these
studies 183 cool white dwarfs with infrared $JHK$ magnitudes measured
on the CIT photometric system. We searched the 2MASS PSC for all white
dwarfs and we recovered the 2MASS $J$, $H$, and $K_S$ magnitudes for
160 stars from our initial CIT photometric sample of 183 objects.

Figure 1 shows the differences in magnitudes between the infrared CIT
and 2MASS photometric systems for the $J$, $H$, and $K/K_S$ filters
for the white dwarfs in our sample. Note that the number of stars in
each panel is different since some stars have not been formally
detected in one or more bands in 2MASS. The size of the error bars in
Figure 1 correspond to the combined quadratic uncertainties of both
data sets, $\sigma=(\sigma_{\rm 2MASS}^2+\sigma_{\rm
CIT}^2)^{1/2}$. For both measurements to be compatible, the error bar
must touch the horizontal dashed line in each panel of Figure 1, which
represents the mean magnitude difference between both data sets, as
determined below.

 \setcounter{figure}{0} \begin{figure}[!h] \centering
 \includegraphics[trim=0 100 0 100, height=11.5cm,angle=0]{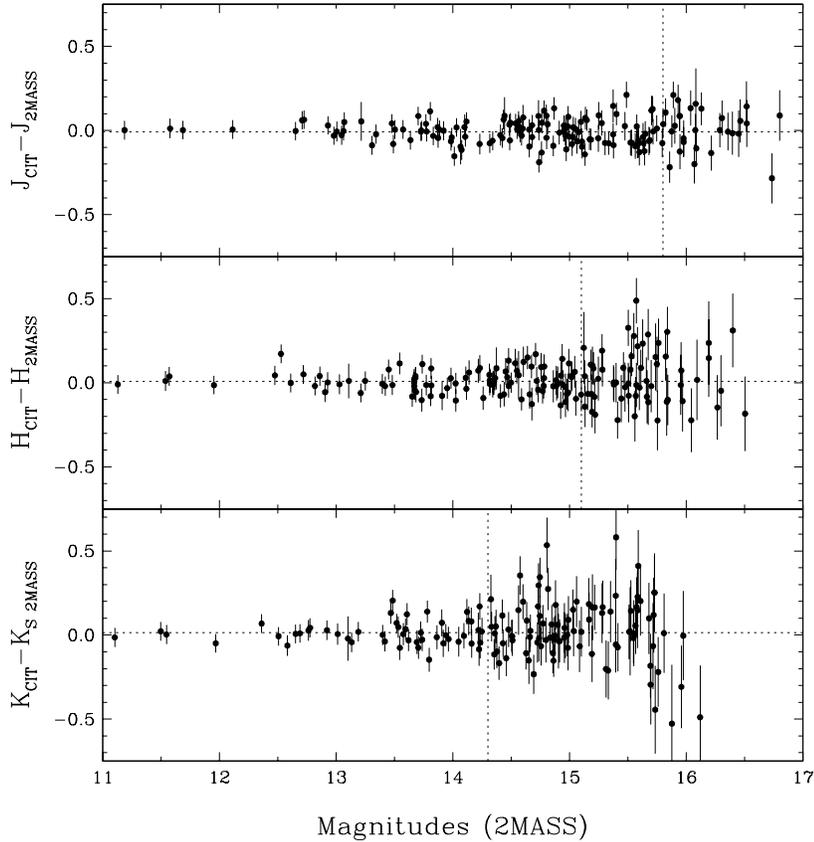}
 \caption{Differences in magnitudes between the infrared CIT and 2MASS
 photometric systems for each individual filter as a function of the
 2MASS magnitude for our common sample of 160 cool white
 dwarfs. Objects located on the left side of the vertical dotted lines
 meet the PSC level~1 requirements (S/N$\,>10$), which correspond to
 $J < 15.8$, $H < 15.1$, and $K_S < 14.3$.}  \end{figure}

We present in Table 1 a statistical comparison of both data sets for
all three bands. The first three lines correspond to the full data set
while the last three lines are restricted to 2MASS magnitudes that
satisfy the level~1 requirements. The second column indicates the
number of stars used for the comparison. The third and fourth columns
represent respectively the mean and the standard deviation of the
magnitude differences for each band. These mean values thus correspond
to the zero point offsets between both photometric systems. We note
that the offsets are typically five times smaller than the average
2MASS uncertainties (fifth column of Table 1) and these could as well
be considered as zero for most practical purposes.

If the uncertainties of both data sets have been properly evaluated,
the average combined quadratic uncertainties, $\langle\sigma\rangle$
(last column of Table 1), should be at least as large as the standard
deviations of the magnitude differences. This is certainly the case
for the level~1 subsample, a result that confirms the reliability of
the 2MASS level~1 photometry. For the complete sample, however, the
$\langle\sigma\rangle$ values are slightly below the standard
deviations. If we assume that the CIT photometric uncertainties have
been properly estimated, which is supported in BRL97 and BLR01 by the
successful fits with white dwarf models, the 2MASS uncertainties might
be slightly underestimated in the case of faint cool white dwarfs near
the survey limit.

\begin{table}[!h]
\caption{Statistical Comparison of CIT and 2MASS Magnitudes}
\begin{center}
 \centering
\begin{tabular}{clcccc}
\hline
\hline
Bandpass&No. of&Mean&Standard &$\langle\sigma_{\rm 2MASS}\rangle$ &
$\langle\sigma\rangle$ \\ $(m _{\rm CIT}$ - $m_{\rm 2MASS})$&Stars
& &Deviation\\
\hline
$J$&159&$-$0.0046&0.0805&0.0502&0.0745\\
$H$&157&$+$0.0180&0.1126&0.0807&0.0997\\
$K/K_S$&143&$+$0.0247&0.1561&0.1096&0.1253\\ $J$ (S/N $>$
10)&130&$-$0.0083&0.0679&0.0409&0.0662\\ $H$ (S/N $>$
10)&97&$+$0.0094&0.0675&0.0502&0.0726\\ $K/K_S$ (S/N $>$ 10) &49
&$+$0.0133&0.0692&0.0466&0.0697\\
\hline
\end{tabular}
\end{center}
\end{table}

\section{White Dwarfs and Low Mass Main Sequence Binaries from 2MASS}

One of the most immediate applications to a large data set of white
dwarf NIR photometry such as 2MASS is to seek infrared excesses due to
cooler companions that are otherwise invisible in the
optical. \citet{wachter03} used a sample of 759 white dwarfs from the
catalog of \citet{ms99} and identified as many as 95 binary candidates
and 15 tentative binary candidates based on the analysis of a
$(J-H,H-K_S)$ two-color diagram built from 2MASS photometry. 

In Figure 2, we compare 2MASS and CIT two-color diagrams for the 143
stars in our presumably single white dwarf sample of \S~2 that have
been detected by 2MASS in all three bandpasses. Using the color
criteria of Wachter et al., illustrated in the left panel of Figure 2,
we find that several binary and tentative binary candidates in both
regions defined by Wachter et al. A comparison with the CIT
photometry, however, reveals that this result can be readily explained
in terms of the larger uncertainties of the 2MASS photometry since
both regions are located $1-2\sigma$ away from the region occupied by
single white dwarfs near the center of the figure. The large amount
of contamination of the binary candidate regions suggests that their
criteria are not stringent enough.

 \begin{figure}[!h] \centering \includegraphics[trim=10 220 30
 200,height=8.0cm,angle=0]{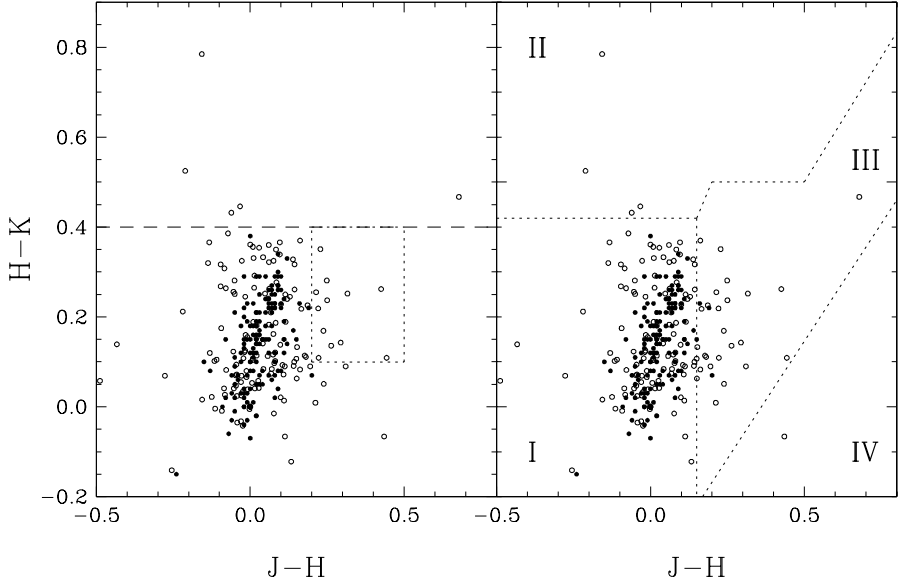} \caption{{\it Left:} $(J-H)$
 vs.~$(H-K/K_S)$ two-color diagrams for 143 cool white dwarfs taken
 from our sample and detected by 2MASS in all three bands. The filled
 and open points correspond to the CIT and 2MASS data, respectively.
 The region above the dashed line and that defined by the dotted
 rectangle correspond to the color criteria defined by
 \citet{wachter03} for selecting binary candidates and tentative
 binary candidates, respectively. {\it Right:} Same as the right panel
 but with the four regions defined by \citet[][see
 text]{wellhouse05}}\end{figure}

\citet{wellhouse05} used a similar two-color diagram approach with a 
sample of 51 magnetic white dwarfs as candidates for potential
pre-cataclysmic variables. They proposed to split the $(J-H,H-K_S)$
two-color diagram into four regions, all illustrated on the right
panel of Figure 2. While they did not find any binary candidates (II),
they identified 10 objects with peculiar colors associated with very
low mass companions or debris (regions III and IV). This represents a
total of 28.6\% of their sample with formal uncertainties with a
possible companion or a disk. From our Figure 2, we find that 21\% of
the white dwarfs from our 2MASS sample of \S~2 would be considered possible
candidates for a companion or a disk, while the CIT data show little
evidence for such infrared excesses. This strongly suggests that the
sample of magnetic white dwarfs studied by Wellhouse et al.~could be
entirely consistent with single stars.

\section{Spitzer Photometry}

To evaluate the reliability of the Spitzer photometry, we use the
observations of \citet{kilic06} who have compared the {\it Spitzer}
4.5 and $8~\mu$m photometric data of 18 cool and bright white dwarfs
with the predictions of model atmospheres. They found that the four
hydrogen atmosphere white dwarfs with $T_{\rm eff}$ lower than 6000~K
show a slight flux depression at $8~\mu$m, while one peculiar object,
the so-called C$_2$H star LHS~1126, suffers from a significant flux
deficit at both 4.5 and $8~\mu$m.

We selected 12 white dwarfs with {\it Spitzer} MIR flux from
\citet{kilic06} which are also in our cool white dwarf sample\footnote{We left out LHS~1126 which indeed exhibits a Spitzer dificit. This peculiar so-called C$_2$H star also show NIR absorption explained by collision-induced opacity. This discrepency may indicate that the collision-induced opacity calculations need to be improved at the high densities encountered in cool white dwarf atmospheres.} of
\S~2. We determine the atmospheric parameters for each star by fitting simultaneously the
average fluxes for the nine photometric bands ($BVRI$, $JHK$/CIT, and
{\it Spitzer} 4.5 and $8~\mu$m). In contrast with the technique used
by Kilic et al., we do not normalize the fluxes at any particular
band, but consider instead the solid angle $\pi (R/D)^2$ a free
parameter. Furthermore, instead of assuming $\log g$$=8.0$ for all
objects, we constrain the $\log g$ value from the trigonometric
parallax measurements. The synthetic fluxes in the MIR are obtained by
integrating our model grid over the {\it Spitzer} IRAC spectral
response curves while the observed fluxes are taken directly from
Table 1 of \citet{kilic06b}. The hydrogen- and helium-rich model
atmospheres used in our analysis are similar to those described in
BLR01 and references therein, except that for the hydrogen-rich models
we are now making use of the more recent collision-induced opacity
calculations and the Hummer-Mihalas occupation probability formalism
for all species in the plasma.

We plot in Figure 3 the ratio of the observed to model fluxes at 4.5
and $8~\mu$m as a function of the derived effective temperature for
the 12 objects. The the agreement between the observed {\it Spitzer}
and model fluxes is very good at all temperatures. In particular,
we do not observe any significant flux deficit at low effective
temperatures as suggested by \citet{kilic06b}. Therefore, we
argue that the results presented in this section demonstrate the
reliability of both the {\it Spitzer} IRAC photometry and our model
atmosphere grid up to $8~\mu$m for studying cool white dwarfs.

\section{Conclusions}

In order to estimate the reliability of the 2MASS photometry for white
dwarf stars, we defined a sample of 160 cool degenerates with $JHK$
magnitudes on the CIT photometric system taken from BRL97 and BLR01,
and compared these values with those obtained from the 2MASS PSC. Our
statistical analysis indicates that, on average, 2MASS uncertainties
are reliable but significant discrepancies are to be expected,
especially for stars near the lower detection threshold. We also
concluded that the search for white dwarf and main-sequence star
binaries based on 2MASS two-color diagrams is greatly limited by the
2MASS uncertainties. We have also shown that the observed MIR
photometry from the {\it Spitzer Space Telescope} agree very well with
our model fluxes.

 \begin{figure}[!h] 
 \centering
\includegraphics[height=13cm,angle=270]{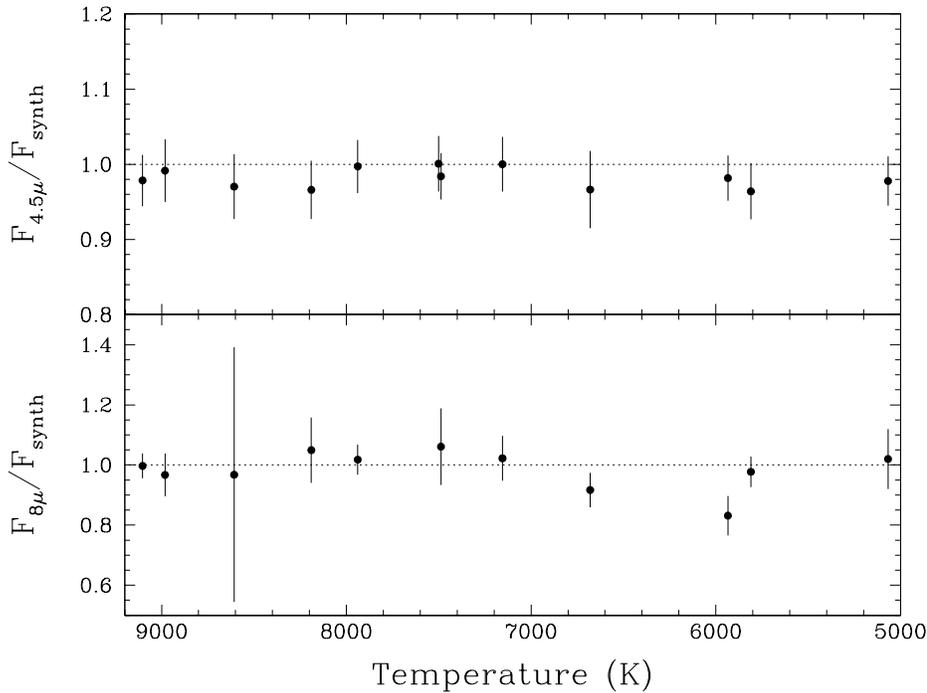}
 \caption{The ratio of observed to predicted {\it Spitzer} fluxes for
 12 objects from the sample of \citet{kilic06b} as a function of
 effective temperature. For Ross 627 (1121$+$216), only the {\it
 Spitzer} $4.5~\mu$m flux is used since the $8~\mu$m flux is affected
 by a nearby star.} \end{figure}

\acknowledgements 
The authors acknowledge the support of the NSERC Canada and the Royal
Astronomical Society. P. Bergeron is a Cottrell Scholar of Research
Corporation. This publication makes use of data products from the Two
Micron All Sky Survey, which is a joint project of the University of
Massachusetts and the Infrared Processing and Analysis
Center/California Institute of Technology, funded by the National
Aeronautics and Space Administration and the National Science
Foundation. This work is based in part on observations made with the
Spitzer Space Telescope, which is operated by the Jet Propulsion
Laboratory, California Institute of Technology under a contract with
NASA.


\end{document}